\documentclass[10pt,british]{article}
\usepackage[T1]{fontenc}
\usepackage[latin9]{inputenc}
\usepackage[letterpaper]{geometry}
\usepackage{babel}

\usepackage{textcomp}
\usepackage{amsthm}
\usepackage{amsmath}
\usepackage{graphicx}
\usepackage{esint}
\usepackage[authoryear]{natbib}
\usepackage[unicode=true, 
 bookmarks=true,bookmarksnumbered=false,bookmarksopen=false,
 breaklinks=false,pdfborder={0 0 1},backref=false,colorlinks=false]
 {hyperref}
\hypersetup{pdftitle={Pinch-off of underwater air bubbles with up--down asymmetry},
 pdfauthor={Daniel C. Herbst}}
\begin{document}

\title{Pinch-off of underwater air bubbles with up--down asymmetry}

\author{Daniel C. Herbst\thanks{herbst@uchicago.edu}}

\maketitle
Physics Department and the James Franck Institute, University of Chicago,
Chicago IL 60637
\begin{abstract}
Topological singularities occur in a broad range of physical systems,
including collapsing stars and pinching fluid interfaces. They are
important for being able to concentrate energy into a small region.
Underwater air bubbles in particular appear in many practical applications,
including new technologies to reduce skin drag on cargo ships. Previous
theories show that just before an air bubble pinches off, the neck
looks like a cylinder at its very smallest point. Unusually, however,
the neck approaches this shape so gradually that the theoretical cylinder
solution is not reached in practice; the singularity spends its entire
lifetime in a transient phase. Therefore, in order to understand the
evolution, we study the transient effects in detail. This paper details
the simulation results of bubbles with initial conditions far from
the cylindrical solution: squat, up--down asymmetric neck shapes,
with imposed vertical flow. We find that the asymmetry is transient:
the neck quickly shifts vertically to become up--down symmetric. Importantly,
we find that the resulting symmetric singularity is a blend of the
initial top and bottom sides, with a weighting factor that is tunable
by adjusting the airflow through the neck. This effect should have
implications for the later stages of evolution, including the generation
of satellite bubbles and the formation of the Worthington jet.
\end{abstract}

\section{Introduction\label{sec:Introduction}}

Non-linear continuous media can produce finite-time singularities,
where a physical quantity nearly blows up in a fixed amount of time.
In practice, the system never becomes fully singular; new forces become
important at small scales and cut off the singularity. Examples of
physical singularities include black hole formation, supernova explosion,
crack propagation, and fluid pinch-off. Here, we study a singularity
that occurs when an underwater air bubble pinches off from its air
source. Despite the small energies involved, the pinch-off produces
a small region of water moving at speeds of 10s of meters per second
\citep{plesset77}. Many of the interesting features of bubble pinch-off
apply to implosion-type singularities in general.

Bubble pinch-off also has practical applications itself. For example,
\citet{mccormick2009drag} discovered that injecting air bubbles under
the hull of a ship reduces skin drag. Only recently has the idea been
implemented on commercial ships, resulting in an overall 5-10\% fuel
savings \citep{ferrante2004physical,mitsubishi2011}. The drag reduction
depends on the size distribution of the bubbles, which in turn depends
on the specifics of pinch-off \citep{longuet-higgins1991,PhysRevLett.98.144503}.
Also, the bubbles can pinch apart in a turbulent flow, generating
acoustic emissions \citep{frizell1987noise} and dispersing into a
distribution of smaller bubbles, which depends on the details of pinch-off
\citep{Lasheras2002247}. The mechanisms controlling bubble pinch-off
are size-independent and also govern the breach of a submarine hull
\citep{ikeda2009}, and the violent whoosh behind an object plunged
into water \citep{PhysRevLett.96.154505}. This paper focuses specifically
on the effects of up--down asymmetry in bubble dynamics. Up--down
asymmetry is present in nearly all bubbles due to the hydrostatic
pressure gradient. In general, the bottom portion of the neck is more
slender than the top due to the hydrostatic pressure. This effect
can be exacerbated if the bubbles are generated in a dynamic stream
of water \citep{oguz93}. In fact, this research has potential implications
for the jet of water that shoots upward after such an impact, called
the Worthington jet after \citet{worthington1897}, \citep{CambridgeJournals:7921450}.
\S\ref{sec:Conclusion} discusses how the initial conditions of the
cavity determine the bubble shape just before the impact that spawns
the Worthington jet. 

\begin{figure}
\begin{centering}
\includegraphics[clip]{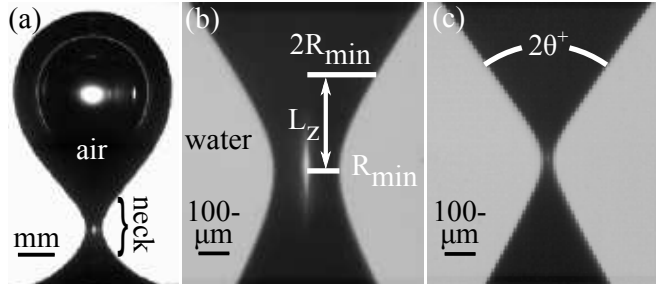}
\par\end{centering}

\caption{\label{fig:experimental_images}Bubble (dark) pinching off from a
nozzle submerged in water (light). Bright spots are optical artefacts.
(a) Initially, the bubble neck shape near the minimum has a generic
quadratic profile. Zooming in (b-c), just before pinch-off the shape
becomes two cones connected at the vertex by a short segment. A characteristic
vertical length scale $L_{z}$ for the neck is the distance from the
minimum to a height where the neck radius is $2R_{\mathrm{min}}$
(b). Even at microscopic scales (c) the neck is relatively squat $(L_{z}/R_{\mathrm{min}}\approx3.2,$
$\theta^{+}\approx28^{\circ}),$ still far from the theoretical end-state
$(L_{z}/R_{\mathrm{min}}\rightarrow\infty$ and $\theta\rightarrow0^{\circ}).$
Neck vibrations eventually cut off the dynamics before the singularity
is reached. (Images courtesy of N. C. Keim and S. R. Nagel).}

\end{figure}

Few people realise that they are creating a singularity when they
blow a bubble into a glass of water using a drinking straw. In experimental
set-ups, the effect can be controlled by precisely pumping air into
a pressurised nozzle (see figure \ref{fig:experimental_images}a).
As the air rises, surface tension competes with buoyancy to create
a quadratic neck shape. Then, the neck becomes unstable and water
rushes in from all sides to close it off. This sets up the regime
we study in this paper, where the inertial flows diverge in speed
where they are closing the shrinking neck, dominating all other forces
and creating a singularity. If this regime were allowed to continue
unaltered, the neck would approach a long-and-slender cylinder, with
purely radial flows \citep{longuet-higgins1991}. Before this occurs,
however, fast air-flows develop in the neck, potentially altering
the collapse scaling \citep{PhysRevLett.95.194501,PhysRevLett.101.214502}.
Eventually, the air even goes supersonic \citep{PhysRevLett.104.024501}.
Just before the airflow effect, some authors including \citet{PhysRevLett.95.194501}
and \citet{PhysRevE.83.056325} measure a small residual asymmetry
that influences the air flow. This is consistent with our results
that the initial asymmetry decays as a power law, but is still measurably
appreciable after at this scale. Finally, \citet{PhysRevLett.97.144503}
found that if the bubble is not quite axisymmetric to begin with,
the deviation actually becomes more pronounced as the neck approaches
pinch-off. The outcome is a side-to-side contact before the average
radius reaches $0,$ cutting off the singularity. These vibrations
were confirmed and studied in detail in later theory \citep{schmidt2009},
experiments \citep{PhysRevE.83.056325,enriquez2011}, and simulations
\citep{Turitsyn2009,lai:102106}.

This singularity is unusual in that it reaches the cut-off point before
the neck shape has a chance to reach its predicted end-state shape,
which is a long cylinder (connecting slender cones on the top and
bottom, see \S2). Most singularities with a unique end-state shape
(\textit{universal} singularities) rapidly approach that shape with
power-law dependence \citep{shi1994}. In those cases, the transient
effects leading up to the universal shape are short-lived and ignorable.
In contrast, the air bubble neck shape evolves very gradually as a
function of the neck radius $R_{\mathrm{min}},$ only going as $\log\left(R_{\mathrm{min}}\right)$\citep{PhysRevLett.95.194501,egg07,PhysRevE.80.036305}.
Indeed, the transition is so slow that on observable length scales
the neck never appears cylindrical, but rather as two relatively squat
cones, smoothly connected (see figure \ref{fig:experimental_images}b-c).
Although technically universal, this singularity spends its lifetime
in the transient phase. Therefore, this paper explores the various
transients that can occur in evolution.

We begin by exploring the effect of highly distorted initial neck
shapes, including necks with large curvature and up--down asymmetry.
We found that the asymmetry is truly transient in every sense: it
vanishes after only several factors of $10$ decrease in the minimum
radius. The neck undergoes this transition not by flexing the cones,
but by shifting the vertical position of the pinch-off. By shifting
instead of readjusting in place, the (new) collapse region rapidly
becomes nearly up--down symmetric. Then, the neck goes through a phase
of being symmetric but also squat, with significant (symmetric) vertical
flows \citep{PhysRevE.84.026313}. Depending on where the neck shifts
vertically, the symmetric squat neck inherits its dynamics from either
its top or bottom side, or a blend of both. Interestingly, we found
that by tuning the initial vertical flow near the neck, we could adjust
the degree of this blending. This effect may explain some of the late-stage
effects seen in experiments that tune the density of the inner gas
\citep{PhysRevLett.101.214502,PhysRevE.83.056325}. When the bubble
is in the formation stage, gas rises through the neck. Denser gasses
impart more momentum to the neck area, giving the neck an initial
vertical velocity component. We predict that this should cause the
neck to inherit its aspect ratio from the top side, which is more
squat. \citet{PhysRevLett.101.214502} do in fact see more squat end-state
neck shapes for higher gas densities (up to their observed critical
density). \citet{PhysRevE.83.056325} observes that the very early
stage air flow during formation sets up a water velocity field, which
then evolves independent of the air during the inertial stage studied
in this paper. He also confirms our observations that the water flow
affects the up--down symmetry and neck position later in the evolution.
Neither experiment, however, attempted to manipulate the initial neck
velocity field purposely, except by altering the gas density. Given
our simulation results, it should be worthwhile to explore more deliberate
ways of controlling the initial airflow through the neck, for example
by blowing the bubble with a top and bottom nozzle simultaneously. 

\S\ref{sec:background} summarises the previous research on the inertial
stage of the bubble singularity, and the need for looking at up--down
asymmetry separately. \S\ref{sec:methods} gives an overview of the
computational methods used to perform the studies. The results are
divided into two sections for ease of reading: \S\ref{sec:ResultsI_shape_asymmetry}
gives a full description of a typical collapse. \S\ref{sec:Results-part-II}
describes how the initial flow field, specifically the vertical velocity
gradient, can be used to tune the details of the final collapse. \S\ref{sec:Conclusion}
concludes with a discussion of how the results could be seen in experiment,
and problems that remain unsolved.

\section{\label{sec:background}Background}

The first stage, described in \S\ref{sec:Introduction}, sets up
the initial conditions for the implosion, and involves the interplay
between surface tension and buoyancy. After this point, surface tension
becomes dwarfed by inertial forces. This is the point where we begin
our simulations, and we assume inviscid flow with only inertial and
pressure forces.

We use the common assumption that the water's velocity field $\mathbf{u}$
is incompressible $(\boldsymbol{\nabla\cdot u}=0),$ irrotational
$(\boldsymbol{\nabla}\times\mathbf{u}=\mathbf{0}),$ and decays to
zero far away. 

We add another assumption that the air in the bubble is dynamically
passive, with a uniform pressure $P_{\mathrm{air}}(t)$ whose value
ensures constant bubble volume in time. Experiments show that after
the neck begins to collapse, airflow is unimportant until the neck
radius reaches roughly 100 \textmu{}m \citep{PhysRevLett.101.214502,PhysRevE.83.056325}.
Around that point, airflow through the neck becomes relevant and our
assumption breaks down. Soon after, neck vibrations dominate. Here,
we will only study what happens up to this point, assuming an axisymmetric
bubble with uniform air pressure.

Being curl-free, the velocity can be described by a scalar potential,
defined as $\boldsymbol{\nabla}\phi\equiv\mathbf{u}.$ Since the exterior
flow is incompressible, $\boldsymbol{\nabla\cdot}\left(\boldsymbol{\nabla}\phi\right)=0$
implies \begin{equation}
\mathbf{\nabla}^{2}\phi=0.\label{eq:Laplace_equation}\end{equation}
That is, Laplace's equation governs the potential in the bulk. This
allows $\mathbf{u}$ to be solved everywhere in the exterior given
$\phi$ on the interface. Equating the relevant stresses on the interface
gives a non-linear differential equation for $\phi$ on the surface
of the bubble, first-order in time and space: \begin{equation}
P_{\mathrm{air}}+\rho\left(\frac{\partial\phi}{\partial t}+\frac{1}{2}\left|\boldsymbol{\nabla}\phi\right|^{2}\right)=0.\label{eq:stressboundary}\end{equation}

where $\rho$ is the water density. The previous two equations determine
the evolution of the velocity field. The velocity field in turn determines
how the interface evolves. Surface elements are advected by the local
velocity on the surface, $S:$

\begin{equation}
\left.\frac{\mathrm{d}\mathbf{x}}{\mathrm{d}t}\right|_{S}=\left.\boldsymbol{\nabla}\phi\right|_{S}.\label{eq:kinematic_condition}\end{equation}

Previously, the solutions to equation \ref{eq:stressboundary} were
only studied carefully in the up--down symmetric, long-and-slender
limit \citep{PhysRevLett.95.194501,egg07,PhysRevE.80.036305}. These
solutions produce elegant, analytic results showing the neck approaching
a cylinder logarithmically slowly as $R_{\mathrm{min}}\rightarrow0$
. There are differences between the theories, but each assumes a (time-dependent)
purely radial flow. This is equivalent to saying that the potential
distribution is proportional to the zeroth order cylindrical harmonic:

\begin{eqnarray}
\phi_{cyl}\left(t\right) & = & -q\left(t\right)\log\left[r/r_{\infty}\left(t\right)\right],\nonumber \\
\mathbf{u}_{cyl}\left(t\right) & = & -q\left(t\right)\boldsymbol{\hat{r}}/r,\label{eq:u_cyl}\end{eqnarray}
 where $r_{\infty}\left(t\right)$ is a length scale related to the
system size, and $q\left(t\right)$ is related to the rate of inflow,
and is determined by equation \ref{eq:stressboundary}. For a perfectly
cylindrical neck with constant $P_{\mathrm{air}}$ and $r_{\infty},$
the reader may verify that $q\left(t\right)=\left[\left(P_{\mathrm{air}}R_{\mathrm{min}}^{2}+\mathrm{const}\right)/\log\left(r_{\infty}/R_{\mathrm{min}}\right)\right]^{\frac{1}{2}}$
satisfies the governing equations. In most cases, the pressure term
is negligible, leaving $q\left(t\right)\sim\left[\log\left(r_{\infty}/R_{\mathrm{min}}\right)\right]^{-\frac{1}{2}},$
which is nearly constant. This defines a flow, $\boldsymbol{u}=-\left[\log\left(r_{\infty}/R_{\mathrm{min}}\right)\right]^{-\frac{1}{2}}r^{-1}\boldsymbol{\hat{r}},$
that evolves the neck gradually towards a cylinder. If $q(t)$ were
instead constant, \citet{longuet-higgins1991} found that any parabola-of-rotation
would generically evolve into a hyperbola-of-rotation. Indeed, at
late stages the bubble neck shape does resemble a hyperbola-of-rotation
by the fact that the neck looks like two smoothly connected cones
(figure \ref{fig:experimental_images}c), which we classify by top
and bottom opening angles, $\theta^{+}$ and $\theta^{-}$ respectively.
For consistency, we actually measure an effective cone angle $\theta_{\mathrm{eff}}$
defined by $2R_{\mathrm{min}}=R_{\mathrm{min}}\sqrt{1+\left(\tan^{2}\theta_{\mathrm{eff}}\right)\left(L_{z}/R_{\mathrm{min}}\right)^{2}}.$
Taking into account the logarithmic dependence of $q$ on $R_{\mathrm{min}},$
the cone opening angles decrease very slowly to $0,$ approaching
a cylinder. Even at the small scale of figure \ref{fig:experimental_images}c,
the cone angle is approximately $28^{\circ}.$ Even under ideal conditions
the neck does not become cylindrical before neck vibrations tear off
the bubble.

\begin{figure}
\centering{}\includegraphics{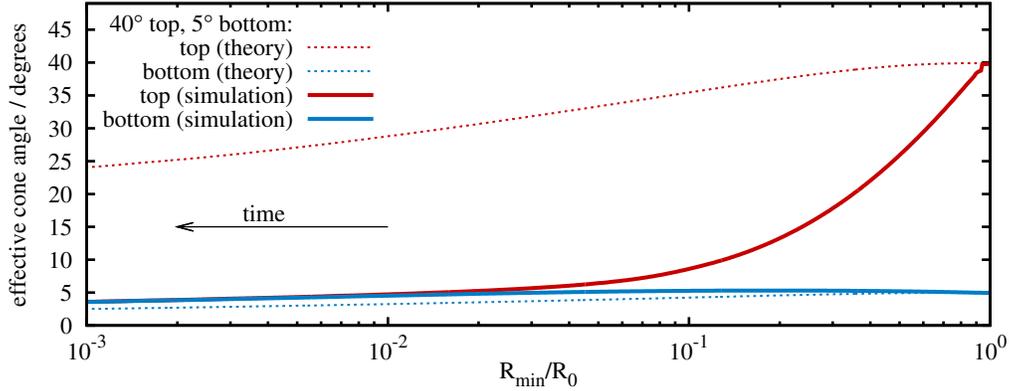}\caption{\label{fig:cone_angle_evolution_vs_cyl}The near-cylindrical theory
of \citet{PhysRevE.80.036305} (dotted lines) fails to properly reproduce
the simulation (solid lines, see \S\ref{sec:methods}) for a neck
shape that is initially up--down asymmetric and squat $(\theta^{+}=40^{\circ}$
and $\theta^{-}=5^{\circ}),$ with initial radial flow. The simulation
reveals that the asymmetry is a short-lived transient, whereas the
theory predicts that the top and bottom sides evolve slowly and separately
to $0^{\circ}$ cone angle.}

\end{figure}

There are two problems with analytic solutions of the form in equation
\ref{eq:u_cyl}. First, they only apply when the neck shape is long-and-slender
and up--down symmetric. Neither assumption holds for real bubble necks
(see figure \ref{fig:experimental_images}). The theory predicts qualitatively
incorrect behaviour for initially up--down asymmetric necks (figure
\ref{fig:cone_angle_evolution_vs_cyl}). In the theory, the neck shape
evolves as a stack of non-interacting cross sections, each one governed
by equation \ref{eq:u_cyl} (or similar). In reality, vertical interaction
plays a large role in the shape evolution. Without vertical flow,
the asymmetry decays logarithmically slowly. With vertical flow, we
find that the asymmetry vanishes as a power law (measured to go as
$R_{\mathrm{min}}^{-1}).$

\begin{figure}
\begin{centering}
\includegraphics{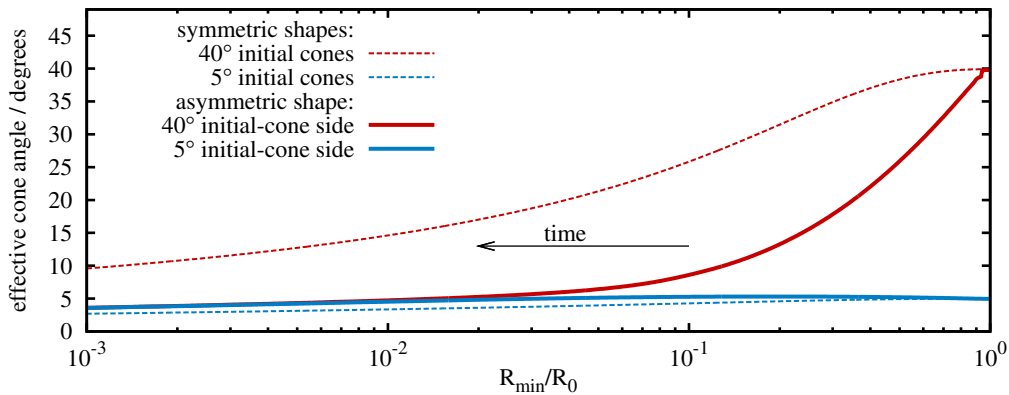}
\par\end{centering}

\caption{\label{fig:Effective-cone-angle}Only when the top and bottom sides
fully interact does our simulation show the up--down asymmetry to
be short-lived (solid lines). For comparison, we also evolved two
symmetric shapes with $\theta^{+}=\theta^{-}=40^{\circ}$ and $\theta^{+}=\theta^{-}=5^{\circ},$
respectively (dashed lines). Taken together, these two simulations
show how the asymmetric shape would evolve if the vertical position
of the neck were fixed (and no flow were allowed through the $z=0$
plane). In that case, the asymmetry decreases very gradually. This
comparison shows that the asymmetry vanishes quickly because the top
and bottom interact. Specifically, the vertical position of the neck
shifts. }
 
\end{figure}

In fact, for up--down asymmetric shapes, the most important effect
comes from the global interaction between the top and bottom sides.
If we artificially forbid flow through the mid-plane, $z=0,$ the
up--down asymmetry no longer vanishes quickly (figure \ref{fig:Effective-cone-angle}
dashed lines). \S\ref{sec:ResultsI_shape_asymmetry} describes the
mechanism in detail. In short, the singularity shifts vertically in
the direction of the smaller cone angle, and in doing so equalises
the cone angles. \S\ref{sec:Results-part-II} shows that this effect
can be manipulated. By introducing a vertical flow to the initial
conditions, the vertical position of the neck can be tuned. In doing
so, the resulting symmetric neck inherits a different blend of characteristics
from the top and bottom sides. This effect should be realisable in
experiments by adjusting the air flow rate through the neck during
the initial set-up phase.

\section{\label{sec:methods}Methods}

To Solve equations \ref{eq:Laplace_equation}-\ref{eq:kinematic_condition},
we use the standard procedure of reformulating the equations as a
boundary value problem, where the bulk differential equation becomes
a surface integral equation. This requires two steps. First, the ordinary
derivative in equation \ref{eq:stressboundary} is replaced by the
co-moving derivative $\frac{D\phi}{Dt}=\frac{\partial\phi}{\partial t}+\boldsymbol{u\cdot\nabla}\phi,$
and will apply to the fluid parcels that constitute the surface. Second,
equation \ref{eq:Laplace_equation} is reformulated using Green's
integral:

\begin{equation}
2\pi\phi\left(\mathbf{x}_{0}\right)=\oint\limits _{\mathbf{x}_{1}\in S}\left(\frac{\left(\mathbf{x}_{1}-\mathbf{x}_{0}\right)\phi\left(\mathbf{x}_{1}\right)}{\left|\mathbf{x}_{1}-\mathbf{x}_{0}\right|^{3}}+\frac{\boldsymbol{\nabla}\phi\left(\mathbf{x}_{1}\right)}{\left|\mathbf{x}_{1}-\mathbf{x}_{0}\right|}\right)\boldsymbol{\cdot}\mathbf{\hat{n}}\mathrm{d}S,\label{eq:green}\end{equation}
 where $\mathbf{x}_{0}$ is a fixed point on $S,$ $\mathbf{x}_{1}$
parameterizes the surface, and $\mathbf{\hat{n}}$ is the unit surface
normal at $\mathbf{x}_{1}$ pointing into the bubble.

With both governing equations defined on the interface, the formulation
reduces to 2 dimensions. We further reduce to 1 dimension by assuming
axisymmetry. At each time step, we use cubic splines to interpolate
the $N(t)$ discrete surface nodes $\mathbf{x}_{i}$. Then, for each
$i\in\left\{ 1\ldots N\right\} $ we allow $\mathbf{x}_{i}$ to take
the place of $\mathbf{x}_{0}$ in equation (\ref{eq:green}). We use
Gaussian quadrature to perform the integral over segments connecting
adjacent spline midpoints $j\in\left\{ 1\ldots N\right\} ,$ with
the first and last segments being only half splines. Thus, the $N$
instances of equation \ref{eq:green} can be schematically written
as $\phi_{i}=\boldsymbol{\mathsf{\mathit{A}}}_{ij}\phi_{j}+\boldsymbol{\mathsf{\mathit{B}}}_{ij}\left(u_{\perp}\right)_{j},$
where $\boldsymbol{\mathsf{\mathit{A}}}$ and $\boldsymbol{\mathsf{\mathit{B}}}$
are $N$-by-$N$ matrices and $u_{\perp}\equiv\boldsymbol{\nabla}\phi\boldsymbol{\cdot}\mathbf{\hat{n}}.$
Therefore, given $\phi$ on the surface we have $u_{\perp}=\boldsymbol{\mathsf{\mathit{B}}}_{ij}^{-1}\left(\phi_{i}-\boldsymbol{\mathsf{\mathit{A}}}_{ij}\phi_{j}\right)$
and $u_{\parallel}=\frac{\mathrm{d}\phi}{\mathrm{d}s},$ where $\mathrm{d}s$
is an infinitesimal arc length. Using $\Delta t\sim R_{\mathrm{min}}/\dot{R}_{\mathrm{min}},$
we advect the N nodes using $\mathbf{u}$ and evolve $\phi$ using
equation \ref{eq:stressboundary}, completing the cycle \citep{pozrikidis,oguz93}.
The pressure $P_{\mathrm{air}}$ can be solved for explicitly after
evolving the shape and before updating $\phi_{i}$ (formula and derivation
available upon request). 

\begin{figure}
\begin{centering}
\includegraphics{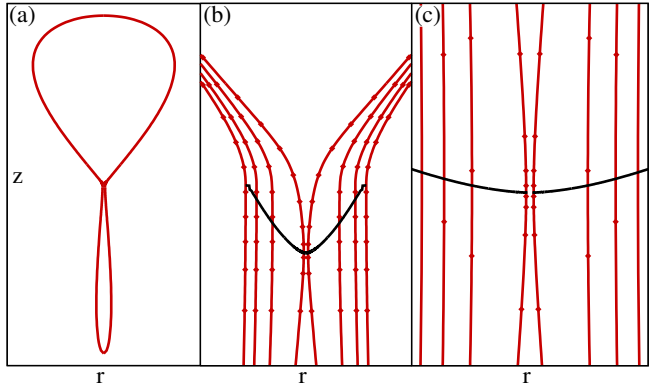}
\par\end{centering}

\caption{\label{fig:shape_evolution} (a) The initial surface is comprised
of an upper half with opening angle $\theta^{+}=40^{\circ},$ and
a lower half with opening angle $\theta^{-}=5^{\circ}.$ The initial
flow is radial. (b) Zooming in $50\times,$ four snapshots show the
neck becoming rapidly more symmetric via a descent of the neck position
(indicated by black curves). In this egregiously asymmetric example,
a second neck actually forms causing a jump in the vertical neck position
(seen as a tiny kink in the black curve in b). (c) Zooming in another
$50\times,$ the simulation evolves the bubble to length scales smaller
than physically possible. Here, the shape is nearly cylindrical and
the neck shift is insignificant. Time steps between curves are: (b)
$0.19\, R_{0}/u_{0r},$ and (c) $4.4\cdot10^{-4}\, R_{0}/u_{0r}.$
Each innermost profile becomes the outermost in the next image. Every
tenth node is shown. }

\end{figure}

In order to accurately resolve the pinch-off dynamics, we found it
important to use a node distribution scheme that maintains a gradual
variation in the spacing between node points and that continually
adds nodes in the neighbourhood of the minimum (see figure \ref{fig:shape_evolution}).
At each moment, if the vertical distance $\Delta z$ between nodes
exceeds a maximum spacing $\Delta z_{\mathrm{max}}(t,\, z)$ in the
region where the neck radius is less than $4R_{\mathrm{min}},$ we
add a new node at the midpoint of that spline. After experimenting
with several functions for $\Delta z_{\mathrm{max}},$ we found $\Delta z_{\mathrm{max}}=\max\left[0.15\min\left(\left|z\right|,L_{z}\right),0.08L_{z}\right]$
allows the simulation to accurately track the dynamics %
\footnote{For the up--down asymmetric case, we define the current height of
the neck minimum as $z_{0},$ $R\left(z_{0}\pm L_{z}^{\pm}\right)=2R_{\mathrm{min}},$
and $\Delta z_{\mathrm{max}}=\max\left[0.15\min\left(\left|z-z_{0}\right|,L_{z}^{+}\right),0.08\min\left(L_{z}^{+},L_{z}^{-}\right)\right]$
for $z\geq z_{0}$ (reverse superscripts for $z<z_{0}).$ %
}. 

We begin the simulation by specifying an interface shape and velocity
field. In general there are an infinite number of each. Here, we examine
generic smooth distributions. To expedite the computation, we do not
prescribe a purely quadratic shape profile at $t=0,$ but instead
use the result from previous studies that a slender quadratic neck
evolves into a hyperbola, and prescribe:

\[
R(z)=\sqrt{\left(R_{0}^{2}+z^{2}\tan^{2}\theta\right)-\left(z^{4}/h^{4}\right)\left(R_{0}^{2}+h^{2}\tan^{2}\theta\right)},\]
where $2\theta$ is the opening angle of the cone, $R_{0}\equiv R_{\mathrm{min}}\left(t=0\right),$
and $h=140R_{0}$ is height from the neck minimum to the end of the
bubble.

To initialise the velocity field, most previous research uses a pure
radial collapse, because it is the counterpart of the cylindrical
end-state solution. We use this as a starting point. Equivalently,
we prescribe a line of point sinks along the z-axis: 

\begin{eqnarray*}
\Phi_{l.s.}\left(r,z\right) & \equiv & \intop_{-L}^{L}\left(r^{2}+\left(z-\zeta\right)^{2}\right)^{-\frac{1}{2}}\mathrm{d}\zeta\\
 & = & \ln\left[\frac{z+L+\sqrt{r^{2}+\left(z+L\right)^{2}}}{z-L+\sqrt{r^{2}+\left(z-L\right)^{2}}}\right].\end{eqnarray*}

We use $L=100,$ much larger than the neck region, to avoid end effects.
For \S\ref{sec:Results-part-II}, we also add a vertical velocity
component to the initial conditions of the neck region. One trivial
possibility is a uniform vertical component. That would be the same,
however, as just boosting the neck region to a new reference frame
$z=u_{0z}t,$ and the pinch-off would be identical, only boosted.
Instead, we use the following form, which consists of a line of sinks
(sources) along the z-axis that vary in strength proportional to height
$\zeta:$

\begin{eqnarray}
\Phi_{z}\left(r,z\right) & \equiv & \intop_{-L}^{L}\zeta\left(r^{2}+\left(z-\zeta\right)^{2}\right)^{-\frac{1}{2}}\mathrm{d}\zeta\nonumber \\
 & = & d_{-}-d_{+}+z\log\left[L-z+d_{-}\right]-z\log\left[-L-z+d_{+}\right],\label{eq:phi_z}\end{eqnarray}

where $d_{\pm}\equiv\sqrt{r^{2}+\left(L\pm z\right)^{2}},$ and the
limit $r\rightarrow0$ is taken for points on the z-axis. This field
acts as a stretched out dipole. The total initial potential field
consists of a weighted sum of the line sink and the dipole field:

\begin{equation}
\Phi_{tot}=\left(\frac{R_{0}d_{0}}{2L}\right)u_{0r}\Phi_{l.s.}+\left(-\frac{2L}{d_{0}}+\log\left[\frac{d_{0}+L}{d_{0}-L}\right]\right)u_{0z}\Phi_{z},\label{eq:phi_line_sink}\end{equation}
where $d_{0}\equiv\sqrt{R_{0}^{2}+L^{2}}.$ The terms in parentheses
are normalisation factors, so that the initial velocity at the neck
minimum $(z=0)$ has radial and vertical components $u_{0r}$ and
$u_{0z},$ respectively. This initial velocity field has the added
benefit of resembling the effect of vertical airflow through the neck
on the surrounding water.

\section{Results\label{sec:Results}}

Our simulations show that the evolution of the neck towards singularity
depends critically on both the initial neck shape and the initial
velocity field. We show the two effects separately. 

\S\ref{sec:ResultsI_shape_asymmetry} details how an up--down asymmetric
shape evolves with a simple initial velocity field: pure radial flow.
The velocity field immediately develops a pressure ring just outside
the neck region, but below the neck minimum itself (in general, towards
the side with smaller cone angle). As the pressure peak moves inward,
it bulges the shape inward, shifting the vertical location of the
neck minimum and making the neck symmetric. The symmetric neck then
progresses as a symmetric collapse, with the upper cone angle adjusted
to match the bottom.

\S\ref{sec:Results-part-II} adds the effect of initial vertical
flow around the neck. A small upward flow is enough to oppose the
neck from shifting downward. After becoming symmetric, the neck is
a blend of the top and bottom cone angles. This gives a possible experimental
method for adjusting the cone angle of the neck at pinch-off.

\begin{figure}
\centering{}\includegraphics{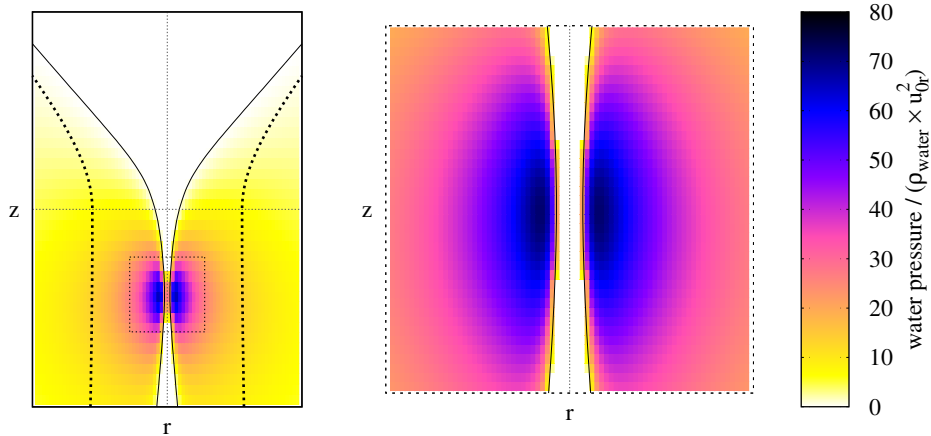}\caption{\label{fig:pressure}The characteristic pressure peak that drives
the singularity develops below neck minimum of the initial shape (original
neck shape: dark dashed curves; original neck minimum: thin horizontal
dotted line), (left). The pressure peak lies just outside the new
neck minimum, with steep gradients between the peak and the surface,
along which the pressure is fixed to be $P_{\mathrm{air}}\left(t\right),$
(right). (Note: this point of evolution corresponds to the centre-most
profile in figure \ref{fig:shape_evolution}b).}

\end{figure}

\subsection{Shape Asymmetry\label{sec:ResultsI_shape_asymmetry}}

We studied many different combinations of initial top and bottom cone
angles $(\theta_{+}$ and $\theta_{-}$ respectively), with each angle
ranging between $5^{\circ}$ and $60^{\circ},$ along with different
initial velocity distributions. The results are qualitatively the
same for any combination of top and bottom cone angles, but the asymmetric
effects become more pronounced for larger asymmetry. In this section,
we use a neutral radial inflow (equation \ref{eq:phi_line_sink} with
$u_{0z}=0$), and an initial $40^{\circ}$ top cone and $5^{\circ}$
bottom cone.

The initial shape consists of the neck, a hyperbola-of-rotation with
the given cone angles, rounded off by large end-caps (figure \ref{fig:shape_evolution}a).
Immediately after the simulation starts, a ring of high pressure develops
in the water surrounding the neck, but slightly below the initial
neck minimum. This causes the shape to bulge in at a height below
the original neck minimum (figure \ref{fig:shape_evolution}b), and
the position of the neck minimum to descend. In our extreme example,
the descent occurs so fast that a new neck actually forms, which appears
as a jump in the neck's vertical position (figure \ref{fig:shape_evolution}b).
Meanwhile, the pressure peak continues to grow in strength and become
more symmetric (figure \ref{fig:pressure}). The neck minimum shifts
downward at roughly a constant rate (lab frame) until pinch-off (figure
\ref{fig:shape_evolution}b). This constant vertical rate is soon
overcome by the large radial implosion flow, and becomes insignificant
at later times (figure \ref{fig:shape_evolution}c). By this time,
the neighbourhood of the neck is completely symmetric, and the system
undergoes the previously-studied symmetric collapse. In this process,
the top side of the shape performs most of the accommodation to match
the bottom side, and the large initial upper angle has little effect
after the transient (figure \ref{fig:Effective-cone-angle}).

\begin{figure}
\centering{}\includegraphics{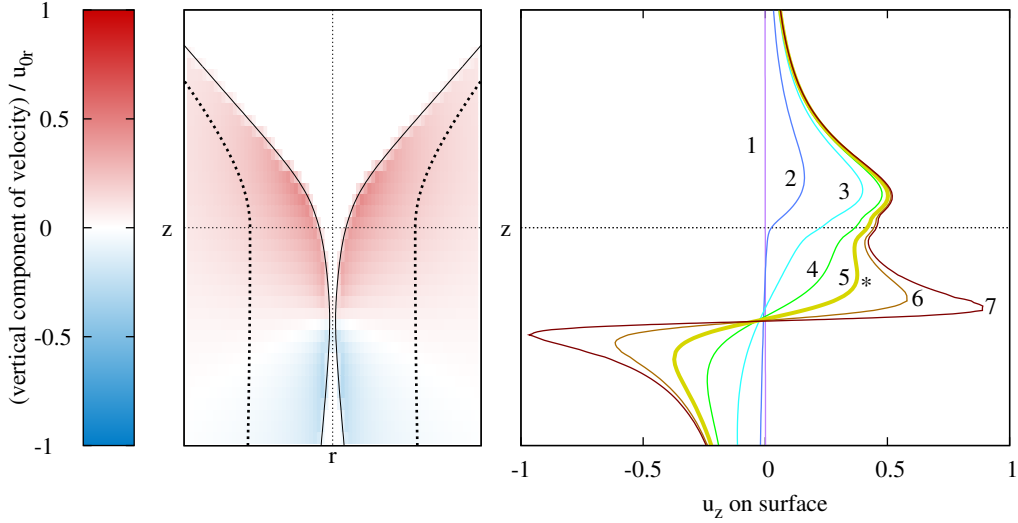}\caption{\label{fig:vertical_velocity_map}(Left) The pressure peak from figure
\ref{fig:pressure} creates a symmetric stretching flow centred below
the original neck minimum, with positive vertical velocity component
$\boldsymbol{u\cdot\hat{z}}$ above the peak and negative below. This
flow leads the shape into a symmetric, vertically-offset collapse.
(Right) Initially, the vertical flow is prescribed with no vertical
flow near the neck minimum (innermost 1). Immediately, the pressure
peak (figure \ref{fig:pressure}) develops, mainly generating positive
vertical velocities on the top side (2). The pressure peak shifts
downward, forming a new, symmetric neck, below the original minimum
(3). Out of this neck, an up--down symmetric straining flow develops
(4-7). Curve 5 (thick) corresponds to the left figure. When the straining
flow begins to develop, it first is only a local maximum on the top
surface ({*}), but continues to grow (6-7), eclipsing the original
peak on the top side, which stagnates and becomes irrelevant. The
straining flow is slightly offset in $u_{z},$ signifying that the
neck continues shifting at a constant rate in the lab frame.}

\end{figure}

The process can be seen in more detail by studying the evolution of
the velocity field, specifically the vertical velocity. Initially,
no vertical flow is present (figure \ref{fig:vertical_velocity_map}
curve 1), but immediately an asymmetric straining flow develops (figure
\ref{fig:vertical_velocity_map} curve 2). Soon afterwords, the origin
of the straining flow shifts downward (figure \ref{fig:vertical_velocity_map}
curve 3). Driven by the pressure ring, the straining flow becomes
symmetric (figure \ref{fig:vertical_velocity_map} curve 4), which
in turn makes the neck shape symmetric. Eventually, the vertical velocity
forms a local maximum in $u_{z}$ on the upper side (figure \ref{fig:vertical_velocity_map}
curve 5, indicated by '{*}'). From then on, the straining flow grows
asymptotically strong, leaving behind only vestiges of the original
top side, which stagnate and become insignificant (figure \ref{fig:vertical_velocity_map}
curves 6-7). The transition from curve 2 to curve 4 shows that the
top part of the straining flow adjusts to match the bottom half through
a downward shift of the straining flow centre. 

\begin{figure}
\centering{}\includegraphics{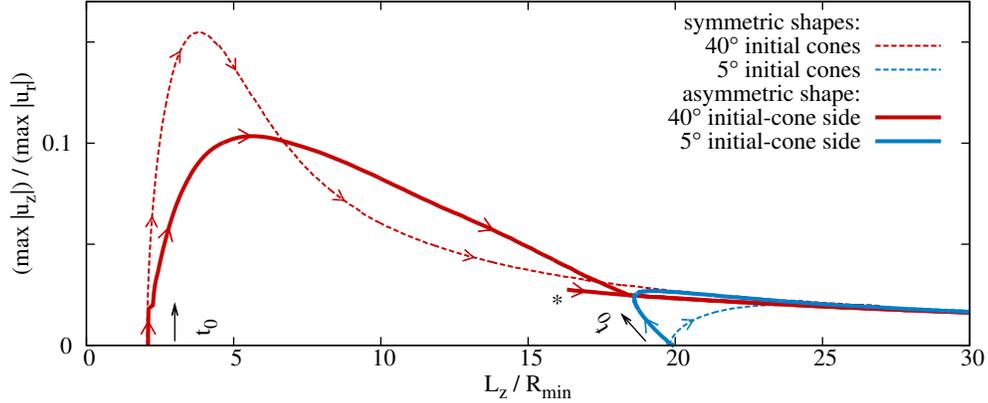}\caption{\label{fig:phase_diagram} The two sides of the asymmetric evolution
(solid curves) evolve drastically differently from their symmetric
counterparts when viewed in phase space (velocity ratio $\left(\max\left|u_{z}\right|/\mathrm{max}\left|u_{r}\right|\right),$
versus the neck aspect ratio $L_{z}/R_{\mathrm{min}}).$ Due to the
strong asymmetry, the two sides see each other and their phase trajectories
attract. The $40^{\circ}$ side is stretched as the neck rolls down,
while the $5^{\circ}$ side is squashed. Meanwhile, the $40^{\circ}$
side is prevented from reaching its full potential vertical velocity.
This is because the singular straining flow is generated on the $5^{\circ}$
side (see figure \ref{fig:vertical_velocity_map}). The newly formed
local maximum vertical velocity (figure \ref{fig:vertical_velocity_map},
'{*}') is represented here also by a '{*}'. This peak eventually prevails
as the global maximum. The two phase curves finally meet, as the neck
region becomes symmetric, then slowly becomes more slender going towards
pinch-off.}

\end{figure}

The shape and velocity field are interdependent, and it is informative
to show how they co-evolve. Figure \ref{fig:phase_diagram} condenses
this information into a phase space. The vertical axis gives the ratio
of the maximum surface vertical velocity component to the maximum
surface radial velocity component (two distinct points). The horizontal
axis gives the shape aspect ratio $L_{z}/R_{\mathrm{min}},$ i.e.,
when the trajectory moves to the right, it corresponds to the neck
shape becoming more long-and-slender. The top and bottom sides each
have separate trajectories, and for comparison, we show each side's
symmetric counterpart (which shows what would happen if no flow were
allowed through the plane $z=0).$ Each trajectory starts on the horizontal
axis, corresponding to the prescribed zero initial vertical velocity,
at a position depending on the initial shape. For example, the $40^{\circ}$
initialised side begins with a corresponding aspect ratio of $L_{z}/R_{0}=\sqrt{3}\cot40^{\circ}=2.06.$
After an initial transient, all the curves coalesce onto a universal
curve \citep{PhysRevE.84.026313} and slowly progress to the universal
form: long-and-slender (large shape aspect ratio) and radial flow
(zero vertical velocity). The progression along the universal curve
has been studied. We focus here on the transient.

Each side of the asymmetric shape differs significantly from its symmetric
counterpart, indicating that the vertical flow through the mid-plane
(interaction between the top and bottom) $z=0$ is important. The
flow through the mid-plane allows the neck to shift until the straining
flow is symmetric. The result is that the trajectories of the top
and bottom sides {}``attract'' one another in phase space. In this
case, the direction of the neck shift is downward. The downward flow
of the neck shift opposes the straining flow on the $40^{\circ}$-initial
top side, and augments the straining flow on the $5^{\circ}$-initial
bottom side. The result is that the peak surface vertical velocity
ratio undershoots its symmetric counterpart on the top side (figure
\ref{fig:phase_diagram} top solid vs. dashed), whereas it overshoots
its symmetric counterpart on the bottom side (figure \ref{fig:phase_diagram}
bottom solid vs. dashed). Similarly, the squat top side shape becomes
more slender and the slender bottom side becomes more squat. In this
way, the two trajectories are brought together in a short time. Some
important features seen in previous plots can also be seen in figure
\ref{fig:phase_diagram}. The local maximum vertical velocity ('{*}'
in figure \ref{fig:vertical_velocity_map}) is shown as a separate
trajectory in figure \ref{fig:phase_diagram} (also indicated by '{*}').
Also, figure \ref{fig:phase_diagram} shows that the bottom side does
not have to detour much in order to meet up with the top side trajectory.
This effect was also seen in figure \ref{fig:Effective-cone-angle}.
In short, since the neck shifts down in this case, the singularity
mostly inherits the characteristics of the bottom side. It is the
top side that adjusts to mirror the bottom side, and the singularity
ignores the fact that the top was initially squat. This effect can
be reversed, as seen in the next section.

\subsection{Tunable cone angle\label{sec:Results-part-II}}

\begin{figure}
\centering{}\includegraphics{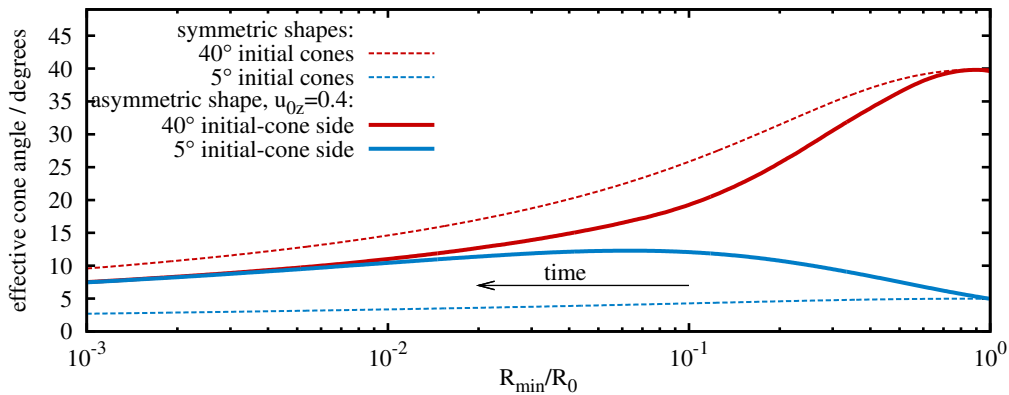}\caption{\label{fig:cone_angle_evolution_with_uz}Adding some vertical flow
$\left(u_{0z}=0.4u_{0r}\right)$ to the initial conditions causes
a significant change later in the evolution. Instead of following
the trajectory of a $5^{\circ}$-initial shape (see figure \ref{fig:Effective-cone-angle}),
the evolution more closely follows the $40^{\circ}$-initial shape. }

\end{figure}

To understand the context for this section, refer back to figure \ref{fig:Effective-cone-angle}.
There, after a short transient, the top and bottom sides (solid lines)
converge to a single trajectory that closely follows the trajectory
of a $5^{\circ}$--symmetric initial shape (lower dashed line). Here,
by adding a vertical velocity to the initial conditions (equation
\ref{eq:phi_line_sink} with non-zero $u_{0z}$), the picture changes.
A prescribed initial vertical velocity 40\% in strength to the initial
radial velocity, $u_{0z}=0.4u_{0r},$ causes the shape to evolve into
a symmetric shape that more closely follows the top-side $40^{\circ}$--symmetric
initial trajectory (figure \ref{fig:cone_angle_evolution_with_uz},
compare to figure \ref{fig:Effective-cone-angle}). The added vertical
velocity opposes the tendency for the neck to shift down in the early
evolution. As a result, when the neck becomes symmetric, it inherits
the characteristics of the top side, rather than the bottom side.
This modest change in initial condition causes a qualitatively different
outcome.

\begin{figure}
\centering{}\includegraphics{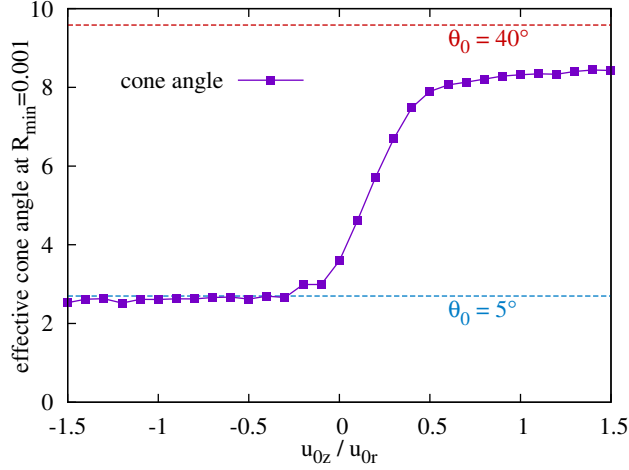}\caption{\label{fig:uz_effects}For neutral and downward initial neck velocities,
the resulting shape after the transient (fixed here as $R_{\mathrm{min}}=0.001R_{0})$
is the same as if the initial shape were $5^{\circ}$--$5^{\circ}$
symmetric (lower line). That is, the upper, $40^{\circ}$ side has
no effect on the late-stage dynamics. For upward initial neck velocities,
i.e. $u_{0z}>0,$ a new effect occurs. The upward flow disrupts the
downward neck shift, and the post-transient neck shape is a blend
of the initial top and bottom sides. Evidently, a modest initial vertical
flow around the neck can tune the properties of the end-state singularity.}

\end{figure}

The effect is not bimodal, but continuous (see figure \ref{fig:uz_effects}).
For $u_{0z}/u_{0r}$ between the range of about $0.0$ to $0.5,$
the resulting trajectory smoothly varies between the bottom-side dynamics
and the top-side dynamics. For initial downward flow $(u_{0z}<0.0),$
the system is insensitive to the exact value; any downward velocity
just results in the singularity inheriting the bottom-side dynamics
(there is no overshoot effect). Interestingly, the dynamics never
fully approach the top-side trajectory, but seem to asymptote slightly
below it. We do not have an explanation for this disparity. 

\begin{figure}
\centering{}\includegraphics{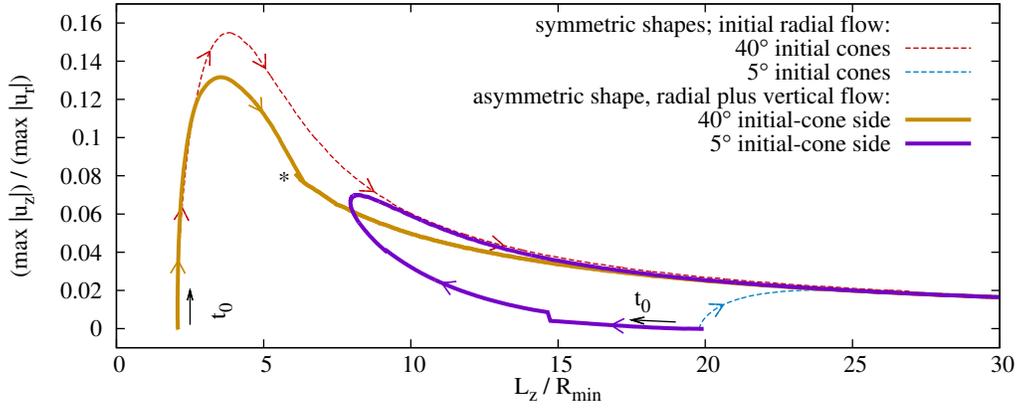}\caption{\label{fig:phase_diagram_uz}Adding a small vertical flow to the initial
conditions has a big consequence on the transient. Here, the initial
shape is the same $\theta_{+}=40^{\circ},$ $\theta_{-}=5^{\circ},$
but now there is an added vertical flow given by equation \ref{eq:phi_line_sink}
with $u_{0z}/u_{0r}=0.4.$ The main effect is that the point where
the two solid lines meet (the point where the shape becomes symmetric)
is significantly further left (corresponding to a larger effective
cone angle). The upward initial velocity opposes the downward neck
shift, and the resulting singularity retains more of the squatness
of the upper side. This effect is tunable by adjusting $u_{0z}/u_{0r}.$ }

\end{figure}

With an added vertical velocity $u_{0z}/u_{0r}=0.4,$ it is the top
side trajectory that dictates the evolution, with the bottom side
going out of its way to accommodate (figure \ref{fig:phase_diagram_uz}).
The point on the phase diagram where the two trajectories merge (becoming
symmetric) is shifted to the left and up, corresponding to a more
squat aspect ratio with larger straining flow. Without the downward
neck shift, the top side evolves almost as if it does not see the
bottom (figure \ref{fig:phase_diagram_uz}, similarity between top
solid and top dashed). The initial vertical flow keeps the neck position
elevated, and the singularity develops within the context of the top
side. The resulting singularity inherits its characteristics from
the top side, nearly ignoring the bottom side completely. A small
initial vertical velocity near the neck can change whether the singularity
inherits from the top side, the bottom side, or a blend of each.

\section{\label{sec:Conclusion}Conclusion}

In general, bubble pinch-off begins with some up--down asymmetry.
We found that the neck adjusts the vertical position of the pinch-off
point to make the neck shape and velocity field symmetric. If the
initial flow field around the neck is purely radial, then the neck
shifts (down) towards the more slender cone. The new neck forms with
a top that emulates the bottom. The resulting singularity depends
only on the initial $\theta^{-}.$ The situation is reversed if a
strong upward velocity is introduced to the neck during the creation
stage. Then, the vertical flow prevents the neck from descending,
and the new neck depends only on $\theta^{+}.$ For intermediate upward
velocities, the shape inherits a blend of the two angles. 

If the singularity were allowed to proceed all the way to $R_{\mathrm{min}}=0,$
this effect would not matter since all symmetric necks, even squat
ones, progress to a cylinder eventually \citep{PhysRevE.84.026313}.
In reality, however, the asymptotic form progresses so slowly that
other dynamics pre-empt the singularity before the neck can become
slender. Therefore, it is the post-transient shape that sets up the
post-pinch-off effects. For example, we expect that tuning the post-transient
cone angle will play a significant role in the later effects, including
neck air flow, azimuthal vibrations, and the Worthington jet, but
more research is needed.

This work was supported by NSF No.~CBET-0967282 and the Keck initiative
for ultra-fast imaging (University of Chicago). The author thanks
Justin Burton, Michelle Driscoll, Nathan Keim, Lipeng Lai, Sidney
Nagel, Thomas Witten, and Wendy Zhang for discussions and suggestions.

\end{document}